\documentclass[conference]{IEEEtran}
\IEEEoverridecommandlockouts

\usepackage{orcidlink}
\usepackage{cite}
\usepackage{amsmath,amssymb,amsfonts}
\usepackage{algorithmic}
\usepackage{graphicx}
\usepackage{textcomp}
\usepackage{xcolor}
\usepackage{comment}
\usepackage{array}
\usepackage{tabularx}
\usepackage{booktabs} 
\usepackage{listings}
\usepackage{mathtools} 
\usepackage[T1]{fontenc}
\usepackage{amsmath,amssymb}
\usepackage{amsfonts}
\usepackage{pifont}
\usepackage{units}
\usepackage{tikz}
\usepackage{regexpatch}
\makeatletter

\def\BibTeX{{\rm B\kern-.05em{\sc i\kern-.025em b}\kern-.08em
    T\kern-.1667em\lower.7ex\hbox{E}\kern-.125emX}}


\begin{document}

\title{Lifecycle Management of Resumés with Decentralized Identifiers and Verifiable Credentials}

\author{\IEEEauthorblockN{Patrick Herbke}
\IEEEauthorblockA{\textit{Service-centric Networking} \\
\textit{Technische Universität Berlin}\\
Berlin, Germany \\
p.herbke@tu-berlin.de}
\and
\IEEEauthorblockN{Anish Sapkota}
\IEEEauthorblockA{\textit{Service-centric Networking} \\
\textit{Technische Universität Berlin}\\
Berlin, Germany \\
anish.sapkota@campus.tu-berlin.de}
\and
\IEEEauthorblockN{Sid Lamichhane}
\IEEEauthorblockA{\textit{Service-centric Networking} \\
\textit{Technische Universität Berlin}\\
Berlin, Germany \\
lamichhane@tu-berlin.de}
}

\maketitle

\begin{abstract}
Trust in applications is crucial, especially for fast and efficient hiring processes. Applicants must present credentials that employers can trust without delays or risk of fraudulent information. This paper introduces a framework for managing digital resumé credentials using Decentralized Identifiers and Verifiable Credentials. We propose a framework for the issuance and verification of Verifiable Credentials in real time without intermediaries. We showcase the integration of the European Blockchain Service Infrastructure as a trust anchor. Furthermore, we demonstrate a streamlined application process, reducing verification times and fostering a reliable credentialing ecosystem across various sectors, including recruitment and professional certification.
\end{abstract}


\section{Introduction} 
Trust is crucial in hiring processes, where the authenticity of resumé credentials drives decision making. Verifying applicants quickly and accurately is a challenge. Traditional verification methods, such as paper-based work certificates, are vulnerable to fraud, making it difficult for companies to ensure the authenticity of resumé credentials. As the "Great Resignation" continues in Europe, the demand for fast and trustworthy hiring processes has become increasingly severe~\cite{ng2023great}.

This study presents a decentralized resumé credential management system that allows applicants to manage their credentials digitally and securely. The proposed solution facilitates the issuance and verification of resumé credentials. By decentralizing the credential management process, our approach streamlines hiring procedures and enhances trust, providing applicants with control over their data while ensuring that employers can depend on the integrity of the credentials they verify.

\textbf{Contribution} 
This paper introduces a Decentralized Application (DApp) designed to manage and verify resumè credentials, using the European Blockchain Service Infrastructure (EBSI)~\cite{EBSIHub2024}. The EBSI was selected for its cross-border trust framework, which ensures secure and efficient verification of credentials throughout Europe. 

Resumé credentials are academic achievements, such as language certificates and job references from employers. Our system integrates Decentralized Identifiers (DID) and Verifiable Credentials (VC) to enable the seamless management, issuance, and verification of resumé credentials~\cite{brunner2020did}. Applicants, as holders, can generate resumès and request VCs from issuers, such as previous employers or educational institutions. Resumé credentials are stored in digital wallets and can be presented to verifiers, such as hiring managers and human resources professionals. Our DApp enables interactions between issuers, holders, and verifiers of resumé credentials. As a result, this paper offers three distinct contributions:

\begin{itemize}
    \item We present a novel resumé credential management framework for creating, issuing, and storing verifiable credentials within a DApp. Our system operates without intermediaries, enabling applicants to manage and employers to verify resumé credentials autonomously.
    \item We implement a real-time communication architecture that facilitates interactions between credential issuers, holders, and verifiers. Our architecture uses a message queuing system to ensure seamless data exchange, enhancing the responsiveness of the credential presentation and verification process.
    \item We leverage the EBSI as a trust anchor for credential verification. The EBSI validates signatures via stored public keys related to DIDs and enhances issuer credibility through its Trusted Issuer Registry (TIR). 
\end{itemize}

\section{Background}
Traditional resumé systems rely on intermediaries, such as universities and employers~\cite{ingold2021resume}. Application processes are plagued by verification delays, fraudulent information, security risks, and transparency issues. Such issues slow the hiring process and compromise the trustworthiness of resumé credentials, making verification difficult for employers~\cite{wanberg2020job}.

DIDs enable individuals to create and control their digital identities without relying on centralized authorities. DIDs are globally unique and cryptographically secure~\cite{sporny2022decentralized}. In the context of resumé credential management, DIDs link VCs to an applicant's identity, ensuring that the identity is both authentic and privately controlled by the individual.

VCs build on the foundation of DIDs by offering a tamper-proof method to issue and verify claims about an individual. VCs are cryptographically signed and allow applicants to share only the necessary information with potential employers. Using VCs in the management of resumés streamlines the verification process, reduces the reliance on third-party verification services, and improves the hiring process~\cite{Sedlmeir2021DigitalIA}.

The EBSI acts as a trust anchor within the proposed DApp of resumé credentials, providing the infrastructure to manage digital credentials across borders. The EBSI supports the verification of resumé VCs by storing public keys of DIDs on-chain and maintaining a TIR that enhances issuer credibility. By integrating the EBSI with DIDs and VCs, our DApp reduces verification delays and strengthens trust among stakeholders in the hiring process.

\section{Related Work}
Digital identity management has advanced with the introduction of DIDs and VCs. These technologies enable self-sovereign identity control, removing the need for centralized authorities and addressing privacy and security concerns in traditional systems. DIDs ensure identity ownership, while VCs facilitate cryptographically secure issuance and verification of claims~\cite{yildiz-interop-tut}.

Blockchain-based resumé credential systems have overcome the inefficiencies of traditional credential management. Kumar et al.~\cite{kumar2020educational} developed a system that uses blockchain to store and verify educational credentials, with the aim of reducing the time required for employers to authenticate these credentials. Similarly, Truver~\cite{taha2020truver} uses Ethereum to create an immutable record of academic achievements. Despite enhanced transparency and security, blockchain-based systems face challenges, such as standardization, limited scalability, privacy concerns, and performance bottlenecks~\cite{10031895, 10031276}. BCERT~\cite{leka2020bcert} is a decentralized academic certificate system that uses blockchain to improve credential security and verification efficiency. BCERT employs Ethereum smart contracts and IPFS for secure storage, using cryptographic hashes and AES encryption to ensure data integrity. The BCERT platform addresses certificate falsification and administrative inefficiencies, offering online and third-party verification in real time.

Our research extends existing blockchain-based resumé credential systems by integrating the EBSI with DIDs and VCs. Our solution addresses the limitations of blockchain-based systems in scalability and privacy, essential features for the verification of cross-border resumé credentials. Unlike previous approaches, our system demonstrates a practical and scalable application of VC-based credential management, providing an alternative to existing methodologies.

\section{Concept}
\begin{figure}[!ht]
    \centering
    \includegraphics[width=0.9\linewidth]{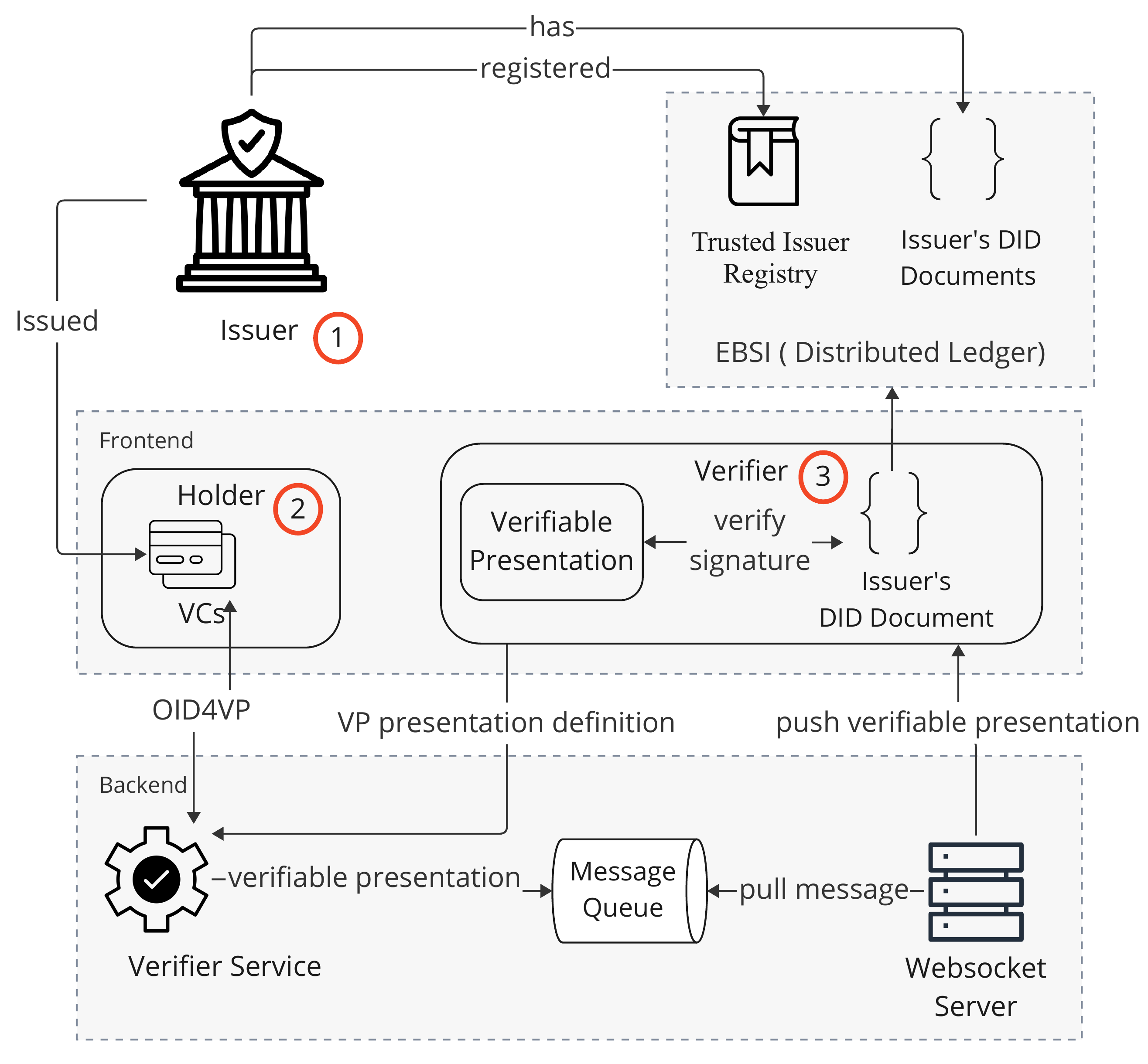}
    \caption{The architecture of the resumé verification dApp integrates issuers, holders, and verifiers, using the EBSI for credential integrity, WebSockets for real-time communication, and a message queue for system resilience and reliability.}
    \label{fig:credential-system-architecture}
\end{figure}

The proposed DApp~\footnote{Resumé DApp: \url{https://github.com/pherbke/seneca/tree/main/usecases/resume}, Accessed 1.9.2024} uses decentralized technologies to manage digital resumé credentials. By integrating DIDs, VCs and the EBSI, our proposed DApp ensures the authenticity and privacy of the resumé credentials. As illustrated in Figure~\ref{fig:credential-system-architecture}, the architecture contains three primary stakeholders: issuers, holders, and verifiers, each contributing to the credentials' lifecycle. 

The flow of resumé credentials, illustrated in Figure~\ref{fig:credential-system-architecture}, begins with the issuer~\textcolor{red}{\textcircled{\raisebox{-0.9pt}{\textcolor{black}{1}}}}, such as a university or employer, which generates VCs for the holder. The issuer signs the VCs with their DID-related private keys, and the corresponding public keys are stored within the EBSI. This setup enables the verification of VCs by checking their cryptographic signatures against the public keys anchored in the EBSI. Additionally, the TIR within the EBSI acts as a trust layer by listing the DIDs of trusted issuers.

The holder~\textcolor{red}{\textcircled{\raisebox{-0.9pt}{\textcolor{black}{2}}}} typically an applicant, can manually create a resumé within the DApp front end, adding positions such as educational achievements or work experience. For each position, the holder can request VCs from the respective issuers. These VCs are stored in the holder's digital wallet. When applying, applicants can initially submit their resumé as a PDF to open vacancies. If interested, the verifier can request a Verifiable Presentation (VP) of specific resumé details. The current implementation allows the presentation of the entire resumé as a single VC. Future work will expand this functionality to enable the presentation of individual VCs with selective disclosure capabilities.

The verifier~\textcolor{red}{\textcircled{\raisebox{-0.9pt}{\textcolor{black}{3}}}} receives the VP and verifies their authenticity by comparing the signature with the public key of the issuer stored on the EBSI. If the VP is verified successfully and the issuer DID is registered in the TIR, the holder's VC is accepted, streamlining the application process. Future enhancements could incorporate immutable registries for trusted schemas and DApps, strengthening the trust framework of the system.

\section{Implementation}
Our proposed DApp integrates components for digital credential management. The Next.js frontend allows holders to create resumés, request VCs, apply for jobs, and submit, as for now a single VP for verification. The frontend interacts with backend services for issuing, storing, and verifying credentials. The backend uses Prisma ORM and PostgreSQL for data management.

The DApp ensures security using the Elliptic Curve Digital Signature Algorithm to sign JSON Web Tokens. For secure communication, the Diffie-Hellman ephemeral static elliptic curve is used to encrypt the credential presentation response. Furthermore, nonces are created using randomly generated numbers to guarantee the uniqueness of cryptographic operations. 

The DApp ensures standardized interactions among issuers, holders, and verifiers using the OID4VC~\footnote{OpenID for Verifiable Credentials:~\url{https://openid.net/sg/openid4vc/}, Accessed 19.9.2024} protocols. A WebSocket server facilitates real-time communication between holders and verifiers, allowing immediate transmission of credential presentations. Redis manages the queueing of messages for interservice communication, maintaining reliability even with sporadic connectivity. The EBSI acts as the trust anchor, storing issuer public keys related to DIDs and DID documents to enable the verification of VCs.

Issuers are legal entities with a~\texttt{did:ebsi}, with the corresponding public key stored on the EBSI blockchain within a DID document. Natural persons, such as applicants, are associated with~\texttt{did:key}, where the public key is the third attribute of~\texttt{did:key:<public\_key>}. Our approach aligns with the requirements of the GDPR~\footnote{GDPR~\url{https://gdpr.eu/}, Accessed 19.9.2024}, ensuring the privacy of citizen data. Furthermore, our DApp integrates the iGrant.io data wallet~\footnote{Grant.io Data Wallet:~\url{https://igrant.io/datawallet.html}, Accessed 19.9.2024}, which stores VCs and supports cryptographic operations to sign and present VCs.

\section{Limitations and Risks}
The proposed DApp presents limitations while providing an innovative approach to resumé management. The dependence of the system on the EBSI introduces a degree of centralization. Issuers must comply with the EBSI requirements to be listed in the TIR, potentially creating a bottleneck and reducing system flexibility. Moreover, reliance on the EBSI constrains the interoperability of the system with other identity ledger technologies, such as cheqd~\footnote{cheqd Website: \url{https://cheqd.io/}, Accessed 19.9.2024}, necessitating refactoring efforts to achieve integration. Another limitation is the potential for malicious exploitation of the issuer's DID. For example, if an issuing institution (e.g., a school) ends operations, the credentials previously issued remain valid. However, if the issuer's DID is not promptly revoked, malicious actors may exploit it. Mitigating these risks requires the timely removal of the issuer from the TIR.

\section{Conclusion}
This research demonstrates the efficacy of decentralized technologies, specifically VCs, DIDs, and the EBSI, in creating a solution to manage digital resumé credentials. By streamlining the hiring process for applicants and employers, our system addresses challenges in credential verification and paves the way for the adoption of decentralized identity management frameworks. Enhancing decentralization with threshold signatures can distribute trust and prevent single points of failure. Future research should ensure interoperability for broader adoption.

\section*{Acknowledgment}
This work was supported by the European Union’s Digital Europe Program (DIGITAL) research and innovation program under grant agreement number 101102743 (TRACE4EU). The authors thank the TRACE4EU project partners for their support.

\bibliographystyle{IEEEtran}
\bibliography{references.bib}

\end{document}